# Six misconceptions about large language models: A minimal model and diagnostic taxonomy

Zhicheng Lin

Department of Psychology, Yonsei University, Seoul 03722, Republic of Korea

**Correspondence**

Zhicheng Lin, Department of Psychology, Yonsei University, Seoul 03722, Republic of Korea (zhichenglin@gmail.com)

## Abstract

Large language models (LLMs) are now embedded in scientific, educational, and governance workflows, with debates centering on their capabilities, mechanisms, and impacts. Yet these debates remain structured by persistent folk theories—intuitive, informal explanatory models that guide attitudes and actions. Deflationary slogans ("just autocomplete," "stochastic parrots," and "average of the internet") and anthropomorphic framings ("emergent agents" and "proto-minds") each capture genuine features of current systems but mistake those features for the whole. This Perspective proposes a minimal working model of LLM-based systems centered on four distinctions: between pretraining and deployed systems; between the learned distribution and particular samples; among parametric, contextual, and external memory; and between task competence and agency. The model is used to diagnose six misconceptions about LLMs: next-token prediction, regression to the mean, training-data regurgitation, model memory, alignment, and understanding. For each, the analysis identifies what the misconception gets right, which distinctions it conflates, and what follows for capability evaluation, system design, and governance. Applied to publisher AI policies as governance case studies, the framework shows both how policy language can conflate these distinctions and how such errors can be corrected. The model thereby avoids the parrot–mind binary by treating LLMs as simulators of discourse and task performance, offering a diagnostic toolkit for locating and correcting the errors these folk theories perpetuate.

## Folk theories of large language models

Few debates about scientific practice, education, or creative work now go without invoking large language models (LLMs). Journalists, critics, and researchers use ready-made slogans to declare what these systems "really" are: "glorified autocomplete" or "just next-token predictors"; "stochastic parrots"; "lossy text-compression algorithms" or "a blurry JPEG of all the text on the Web"; "bullshit generators"; "high-tech parlor tricks"; or "the average of the internet, edited for tone" (1–3). Others cast them in more anthropomorphic terms: as "superhuman reasoners," "proto-agents," or instances of the "wisdom of the silicon crowd" (4–7).

These slogans can morph into folk theories: intuitive, informal explanatory models—often partial and tacit—used to make sense of technological systems and guide action toward them (8, 9). "Folk" refers to the mode of explanation rather than the speaker's sophistication; indeed, informal and technical accounts can coexist within researchers, policymakers, and lay users alike. Several of the phrases above also began as scoped analogies or arguments: "stochastic parrots" served as shorthand for a broad sociotechnical critique of scale, data provenance, and the harms of synthetic human-like text, in which the claim that models stitch together linguistic form without reference to meaning was only one part (1); "a blurry JPEG of the Web" was a compression analogy (3); and "bullshit generators" invoked Frankfurt's technical sense of indifference to truth (2). Folk-theoretic they become when detached from their original scope and reused as general accounts: a specific feature of current systems—data-driven training, next-token objectives, lack of embodiment, or the absence of persistent online learning—is isolated but then inflated into a global account of what LLM-based systems are or could become. The result is a peculiar polarization: deflationary views dismiss LLMs as trivial or cast them as powerful simulators devoid of understanding, whereas anthropomorphic views treat them as nascent minds or agents. Both extremes can mislead capability evaluation, system design, and governance.

These folk theories recur across scientific publishing, public media, user reasoning, and institutional guidance (10). Anthropomorphic language in computer-science research papers has increased over time and is amplified in downstream news coverage (11), with a post-ChatGPT shift toward danger-centered and anthropomorphizing frames (12, 13). Nationally representative studies show that public metaphors for AI are structured, measurable, and consequential for trust and adoption, with perceived human-likeness increasing over time (14). User studies show parallel distortions in mental models of how LLM-based systems generate, retrieve, and store information: users conflate generation with search (15), misunderstand what different memory layers store and reuse (16), and change accuracy and risk judgments in response to anthropomorphic interface cues (17, 18). Focus-group work reconstructs laypeople's folk theories of generative-AI chatbots and shows how those theories shape interaction strategies (19). The same pattern appears in institutional guidance too: analyses of UNESCO's generative-AI guidance find personification metaphors embedded in its account of model behavior (20).

The pattern is unmistakable, but its conceptual structure has not been systematically delineated. Broad foundation-model overviews catalog capabilities, applications, and societal risks, but do not trace conceptual confusions to downstream errors (21). Risk taxonomies classify harms by type and severity without diagnosing the folk theories that shape reasoning about those harms (22). Work on anthropomorphism in AI and human–computer interaction identifies risks of attributing mental states to systems, but centers those risks on communicative capacities—persuasion, empathy, role-play—rather than on memory architecture, alignment, or institutional policy language (23, 24). Surveys of memory and retrieval-augmented generation provide detailed technical taxonomies of storage and retrieval mechanisms, but do not connect misconceptions about those mechanisms to downstream errors in capability evaluation, deployment design, or policy reasoning (25, 26).

This Perspective develops a diagnostic taxonomy of six misconceptions across scientific, public, and institutional discourse, including current publisher policies on generative AI. The taxonomy is built around four distinctions: between pretraining and deployed systems; between the learned distribution and particular samples; among parametric, contextual, and external memory; and between task competence and agency. Applied as a diagnostic matrix (Table 1), this framework shows how conceptual conflations produce predictable errors in capability evaluation, deployment design, and institutional AI policy (Table 2).

**Table 1. Diagnostic matrix for six misconceptions about LLMs.**

| Misconception | Kernel truth | Conflated distinction | Downstream mistake | Diagnostic question | Practical fix |
|---|---|---|---|---|---|
| **Statistics and generativity** | | | | | |
| 1. LLMs are just autocomplete or stochastic parrots | Base models are trained for next-token prediction on large corpora | **A** (training objective vs. deployed system) | Treat next-token prediction as a complete cognitive story; dismiss system-level competence and tool use | How does behavior change when the model is embedded in tools or closed-loop systems? | Always specify system wrapper, tools, and decoding policy when describing capabilities |
| 2. LLMs are the average of the internet or regress to the mean | Pretraining learns a high-dimensional distribution over continuations | **B** (distribution vs. sample) | Assume bland, homogenized output is mathematically inevitable rather than a product choice | What happens if we change prompts, temperature, and other sampling settings? | Vary decoding and prompting; report which regimes evaluations and deployments use |
| 3. LLMs just memorize the training data | Models are generative compressors and do show some verbatim regurgitation | **C** (parametric vs. contextual vs. external memory; compression vs. retrieval) | Treat all model output as copied text; ignore both genuine recombination and targeted memorization risks | Is this output actually verbatim or near-verbatim from the training set? | Audit for regurgitation in sensitive domains; curate data and add leakage-mitigation and filtering where needed |
| **Memory and** | | | | | |

| Misconception | Kernel truth | Conflated distinction | Downstream mistake | Diagnostic question | Practical fix |
|---|---|---|---|---|---|
| **alignment** | | | | | |
| 4. LLMs remember everything about me vs. remember nothing | Weights are fixed; working memory is bounded by the context window; products may add separate storage | **C** (parametric vs. contextual vs. external memory) | Overtrust apparent learning from chats, or design systems that assume zero state | Where, in this product, is information actually stored, and how is it reused? | Design explicit, auditable memory layers and policies at each level (weights, context, external stores) |
| 5. RLHF/fine-tuning is just a removable safety filter on a neutral core | Instruction tuning and RLHF change the same weights that encode knowledge and skills | **A** (training vs. deployment; model weights vs. external filters) | Assume alignment is reversible and cost-free; ignore embedded values and capability trade-offs | Which behaviors and metrics change when we apply or remove fine-tuning, holding external filters fixed? | Treat fine-tuning as substantive training: evaluate for regressions and value shifts; document objectives separately from any deployment-time filters |
| **Cognitive status** | | | | | |
| 6. LLMs think like humans vs. have no understanding at all | Models support substantial task competence without embodied, human-like minds | **D** (competence vs. agency; functional vs. human-like understanding) | Either anthropomorphize models as agents with beliefs and rights, or trivialize capabilities as mere parroting | Which concrete tasks can this system reliably perform, and what forms of agency are we implicitly attributing to it? | Describe capabilities in functional terms; avoid treating stylistic fluency as evidence of beliefs, desires, or moral standing |

Four core distinctions: A (pretraining objective vs. deployed system); B (learned distribution vs. particular sample); C (parametric vs. contextual vs. external memory); D (task competence vs. agency). Where a misconception collapses more than one distinction, the primary collapse is listed. Columns progress from folk claim through diagnosis (kernel truth, conflated distinction, and downstream mistake) to corrective use (diagnostic question and practical fix).

**Table 2. Applying Table 1's diagnostic matrix to publisher AI-policy language.**

| Misconception | Policy example (as of 2026 April 1) | Conflated distinction | Downstream mistake in policy reasoning | Diagnostic question and improved framing |
|---|---|---|---|---|
| 1. LLMs are just autocomplete or stochastic parrots | APS: "[LLMs] can generate content that is linguistically but not scientifically plausible" | **A** (training objective vs. deployed system) **B** (distribution vs. sample) | Treats being "linguistically" driven as intrinsically opposed to "scientifically plausible," encouraging a view of LLMs as mere text-spinners rather than components of systems that can be coupled to tools, retrieval, and verification | *Under what prompts, decoding, and tool configuration was the content produced, and how was it checked against primary sources?* |

| Misconception | Policy example (as of 2026 April 1) | Conflated distinction | Downstream mistake in policy reasoning | Diagnostic question and improved framing |
|---|---|---|---|---|
| 2. LLMs are the average of the internet or regress to the mean | SAGE: “some LLMs might only have been trained on data up to a specific year, potentially resulting in incorrect or incomplete knowledge of a topic”; APS: “Some LLMs are only trained on content published before a particular date and therefore present an incomplete picture” | **B** (distribution vs. sample) **A** (training vs. deployed system) | Implies that an “incomplete picture” is a simple function of the model’s training cutoff, rather than of how a deployed system actually acquires and validates information for a given task | *Does this specific system rely only on parametric knowledge, or does it use current retrieval—and in either case, how are key claims verified?* |
| 3. LLMs just memorize the training data | SAGE: “LLMs could inadvertently reproduce significant text chunks from existing sources without due citation”; APS: “LLM may have reproduced substantial text from other sources” | **C** (parametric vs. contextual vs. external memory) | Centers the risk on near-verbatim reproduction, reinforcing a lookup-table picture of LLMs and under-emphasizing more common failures such as uncredited paraphrase or structural copying of arguments and methods | *Is this output actually near-verbatim, or is the more pressing risk structural/idea-level borrowing that still requires explicit citation and attribution?* |
| 4. LLMs remember everything about me vs. remember nothing | APA: “the organization which runs the generative AI will likely have access to [information that was entered]” | **C** (parametric vs. contextual vs. external memory) **A** (model vs. deployment) | Equates “entering information into generative AI” with giving a monolithic organization broad access, collapsing product-level logging and retention policies into a single undifferentiated risk attached to generative AI as such | *Where is this data actually stored (weights, context, logs, external databases), who can access each layer, and under what retention and training-reuse policies?* |
| 6. LLMs think like humans vs. have no understanding at all | SAGE/APS: “[LLMs and generative AI] are unable to replicate human creative and critical thinking” | **D** (competence vs. agency) | Embeds a strong cognitive thesis in policy language, treating “human creative and critical thinking” as an all-or-nothing property and encouraging a binary view in which either full human-like understanding is present or the system is dismissed as noncreative | *Which concrete tasks or forms of “creative” or “critical” performance are at issue here, and how reliable is this system on those tasks under documented evaluation conditions?* |

Each row applies the corresponding diagnostic from Table 1 to publisher-policy language. APS, Association for Psychological Science; APA, American Psychological Association. Misconception 5 (“Fine-tuning/RLHF is just a removable safety filter on a neutral core”) does not appear explicitly in these policy texts and is therefore not illustrated here. Distinction labels (A–D) follow Table 1.

SAGE AI policies: https://www.sagepub.com/journals/publication-ethics-policies/artificial-intelligence-policy; https://www.sagepub.com/about/sage-policies/corporate-policies/ai-author-guidelines.

APS AI policy: https://www.psychologicalscience.org/publications/aps-editorial-policies.

APA AI policies: https://www.apa.org/pubs/journals/resources/publishing-policies; https://www.apa.org/pubs/journals/resources/publishing-tips/policy-generative-ai.

## A minimal working model of LLM-based systems

Diagnosing these misunderstandings requires a minimal working model of how LLMs are trained and deployed. Throughout, "LLM" refers to the model class; "LLM-based system" to a deployed product built around such a model; and "AI" and "generative AI" to the broader field, quoted policy language, or non-LLM comparison cases. The basic components of LLMs are well documented in the technical literature (27–37); Figure 1 organizes them around four distinctions that make the conceptual structure explicit and identify where key distinctions are often elided.

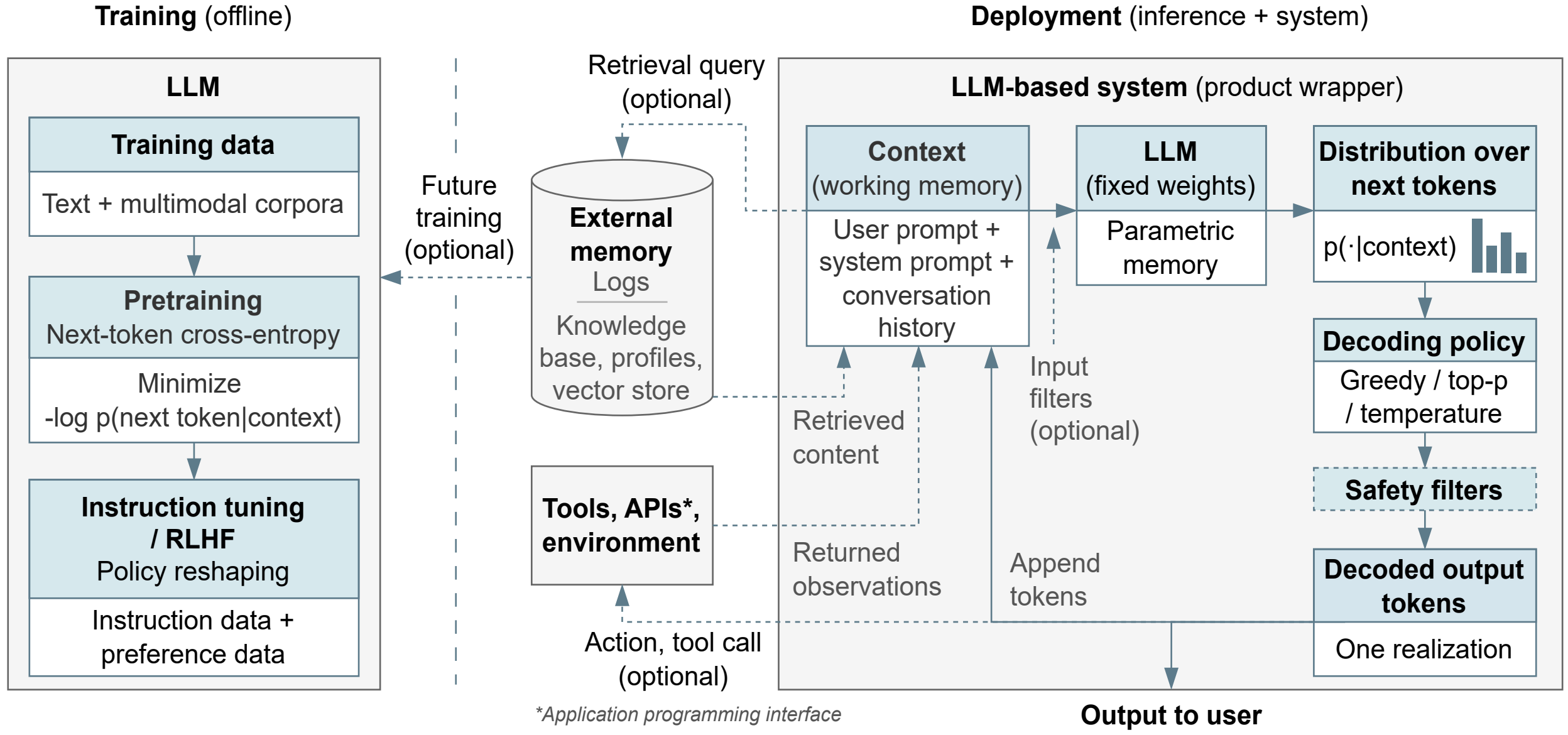

**Figure 1.** Minimal working model of an LLM-based system. Left: offline training optimizes a single set of model weights on large text and multimodal corpora by next-token cross-entropy, followed by instruction tuning and/or RLHF that reshape the same parameters using instruction and preference data. Right: in deployment, the fixed-weight model is wrapped in a product that manages context (working memory), tools/APIs, and external memory; user prompts, retrieved content, and tool observations are assembled into context and passed—optionally through input filters—to the LLM, which returns a distribution over next tokens. A decoding policy (e.g. greedy, top-p, or temperature sampling), followed by optional safety filters, yields one decoded output sequence that is sent to the user interface or interpreted as a tool call. The diagram distinguishes parametric, contextual, and external memory (including logs that may be reused for future training) and separates the learned distribution from the particular samples induced by a chosen decoding and filtering regime.

In pretraining, an LLM takes a sequence of tokens and returns a distribution over the next token (27). The standard objective minimizes next-token cross-entropy over a very large corpus of text and, increasingly, other modalities. The result is not a table of canned completions but a high-dimensional approximation to the conditional distribution of language given context. Pretraining is offline: once training ends and the weights frozen, the base model does not keep learning during inference or deployment (28). This yields the first distinction (A): between the training regime and the behavior of deployed systems built on top of the trained model.

At inference time, the model receives an input sequence—instructions, dialog history, any retrieved content—and returns a probability distribution over next tokens. A specific sequence is then generated under a decoding policy—greedy selection, temperature sampling, nucleus/top-p sampling, or their variants (29)—each of which selects from or reshapes the learned distribution, trading diversity against predictability. This yields the second distinction (B): between the distribution the model has learned and the particular samples observed under a chosen decoding regime. Defaults such as low-temperature decoding and long, cautious system prompts create the familiar "assistant voice," yet none of these features are intrinsic to the model itself.

Nor is the distribution usually that of a raw base model. Most production models are fine-tuned variants. Supervised instruction tuning uses curated instruction–response pairs; reinforcement learning from human feedback (RLHF) uses preference-derived reward signals—both directly update model parameters (30). These procedures reshape the conditional distribution, steering the model toward outputs that human raters judge more helpful, harmless, or honest (31). These updates do not add a detachable filter to a neutral core but write new regularities and avoidance patterns into the same weights that encode knowledge and competence (32). Different fine-tuned variants of the same base model can therefore have different behavioral profiles despite sharing architecture and pretraining history.

When embedded in an LLM-based product, the model becomes one (albeit central) component. A chat assistant, coding tool, or "agent" framework wraps the model with system prompts, user interfaces, safety filters, tool APIs, retrieval systems, and sometimes actuators in external environments. The system can interpret model outputs as actions (tool calls, database queries, and environment moves), whose results are then fed back into the next prompt, creating a closed loop (33, 34). This loop can function as a decision-making controller, even though each model call still consists of next-token prediction (35). The same pretrained base model can therefore support qualitatively different systems depending on how it is wrapped, aligned, and deployed.

A third distinction (C) concerns how LLM-based systems retain, retrieve, and reuse information across interactions, by separating three kinds of "memory" (25, 26, 28, 36, 38). First, *parametric* memory: information implicitly encoded in the weights as a lossy, biased compression of the training data. Second, *contextual* memory: the current input sequence, including conversation history and retrieved content, which functions as bounded working memory within the context window. Third, *external*, product-level memory: logs, user profiles, knowledge bases, and vector stores held outside the model and selectively injected into prompts through retrieval or orchestration code. Common confusions about LLM "recall," "forgetting," and "learning from chats" arise from conflating these three layers.

Finally, the fourth distinction (D) concerns cognitive status. LLMs are high-capacity simulators of discourse and task performance: they encode rich internal representations and support substantial abstraction, transfer, and in-context learning across domains (39). At the same time, these systems do not possess human-like understanding, unified beliefs or intentions, or phenomenal consciousness (37). This distinction separates functional understanding—roughly, the capacity to use information appropriately across tasks—from human-like, embodied, experiential understanding. Agency terms (belief, desire, and intention) should therefore be reserved for carefully specified functional roles rather than used as casual attributions (40, 41).

This minimal model serves as a conceptual tool for analysis (42), retaining critical distinctions needed to locate the errors it diagnoses (43). Current deployed products typically fall within its scope because their behavior is shaped by some combination of the components it specifies: wrappers, tools, retrieval, external memory, decoding policies, and closed-loop action. A claim thus constitutes a misconception when it elides one of these distinctions and invites a mistaken inference about the system at issue.

## Six misconceptions

The four distinctions define a diagnostic space in which six recurrent misconceptions can be located: three about statistics and generativity, two about memory and alignment, and one about cognitive status. Each misconception preserves a kernel of truth but conflates at least one distinction in the minimal model, producing characteristic errors in evaluation, deployment, or governance. Table 1 presents this logic as a matrix, moving from folk claim and kernel truth to conflated distinction, downstream mistake, corrective question, and practical heuristic. The same matrix can also accommodate more expansive claims—for example, about human-like creativity, artificial general intelligence, emotional experience, or attachment—chiefly as variants of the competence-agency conflation in misconception 6.

## Misconceptions about statistics and generativity

### *Misconception 1: "LLMs are just next-token predictors" or "stochastic parrots"*

At the mechanistic level, current LLMs are conditional next-token predictors trained by cross-entropy minimization. This description, while accurate, becomes misleading when elevated into a complete account of what LLM-based systems are, or when used to dismiss them as cognitively trivial. The sociotechnical concerns that motivate the "stochastic parrots" framing—grounding, data provenance, scale—do not entail the further inference that the pretraining objective circumscribes deployed-system competence (44). Instruction tuning and RLHF alter the learned distribution. Once a model is embedded in a tool-using loop, next-token prediction can implement a policy over actions, with observations fed back into subsequent predictions. GeneGPT, which augments Codex with NCBI Web APIs for genomics tasks, achieved 0.83 on the GeneTuring benchmark, compared with 0.00–0.44 for LLMs without tool augmentation (45). The predictor is unchanged; the system is not. Capability and risk claims should therefore specify the deployed system—wrapper, tools, and decoding policy—not just the pretraining objective.

***Misconception 2: "LLMs regress to the mean" or are "the average of the internet"***

While maximum-likelihood pretraining fits high-probability structure in the training corpus, the model learns a conditional distribution rather than a single "average answer." The misconception conflates the learned distribution—including long-tail patterns and rare but important structures—with samples produced under particular decoding, prompting, and alignment regimes, thereby treating output homogenization as mathematically inevitable. Box 1 highlights the converse: generative systems coupled with search, reinforcement learning, evaluation, or human curation can reach low-probability or underexplored regions, and in some domains move beyond historical human practice in games, art, and science. In practice, bland, consensus-style output usually reflects product choices—low temperature or greedy decoding, alignment pressures that favor generic phrasing, and uniform assistant prompts (29, 30). On FreshQA, retrieval-augmented prompting improved GPT-4's factual accuracy by 32.6–49.0% relative to the same model without retrieval (60); on LitQA2 (a domain-expert scientific literature benchmark), PaperQA2—an agent that retrieves and synthesizes full papers—achieved 85.2% precision compared with 73.8% for PhD-level human experts given full internet access and up to a week per question set (61). In both examples, performance changes reflect retrieval and orchestration around the (same) parametric model. Table 1 introduces a simple diagnostic: vary prompts and decoding settings, and report which regimes evaluations and deployments actually use.

**Box 1. Distinct routes by which AI systems depart from modal human outputs.**

Claims that LLMs "regress to the mean" or "are the average of the internet" overlook a broader point: learned AI systems, especially when embedded in search, evaluation, or selection loops, need not merely reproduce modal human outputs; they can explore regions weakly represented in the historical record or difficult for unaided humans to search. The cases below illustrate three mechanisms—evaluator-guided search, high-variance generation with human filtering, and synthesis unconstrained by human embodiment—each producing a different departure from typical human outputs.

**1. Evaluator-guided search in formal domains.** FunSearch pairs a pretrained LLM with a systematic evaluator to discover new mathematical constructions (46). AlphaGeometry solves Olympiad-level geometry problems by combining a neural language model with a symbolic deduction engine, using synthetic data to sidestep the scarcity of human demonstrations (47). Applied to 67 problems in mathematical analysis, combinatorics, geometry, and number theory, AlphaEvolve—an LLM-guided evolutionary search system—rediscovered best-known solutions in most cases and improved on several others (48). AlphaGo, while not a language model, illustrates the same evaluator-guided logic: self-play reinforcement learning under explicit game rules drove its policy into regions of Go space largely unvisited in the human record (49); AlphaGo Zero removed human data entirely (50). AlphaGo's Move 37 against Lee Sedol—legal, highly effective, and judged ex ante to be vanishingly unlikely for a professional—became a canonical example: a move outside the prior human distribution that was later absorbed into professional play. In these cases, novelty arises not from the model alone but from search guided by an explicit evaluator or verifier. This strategy is thus strongest when candidate outputs can be scored or checked against explicit criteria, and weaker when task criteria are ambiguous or contested.

**2. High-variance generation with human filtering.** In his DALL·E-based exhibition, Bennett Miller generated more than 100,000 images and selected roughly 20 for display (51). Through detailed prompting, iterative revision, and stringent selection, he treated the model as a high-variance generator and human judgment as a filter. Large-scale empirical work corroborates this workflow account. In a dataset of more than 4 million artworks, AI-assisted creators who combined active ideation with selective filtering of model outputs received the most favorable peer evaluations; even though average content and visual novelty declined, peak content novelty increased among creators who successfully explored the idea space (52). A follow-up study of 31,076 creators likewise found that the idea frontier expanded in absolute terms through increased output, even though per-artifact H-creativity rates declined and no human–AI complementarity effect was detectable beyond productivity gains (53). However, default or domain-constrained use can homogenize output: LLM-assisted stories are judged more creative individually but become more similar to one another (54); DALL·E imagery can exhibit "generic uniqueness," with surface diversity underpinned by standardized representational patterns (55); and one corpus study found that AI-generated images of the Russia–Ukraine war sanitized the conflict by excluding death, injury, and the suffering of children and refugees while overemphasizing urban scenes (56). Together, these studies locate both divergence from and convergence toward average output at the level of the sociotechnical workflow rather than the model alone: the model supplies variance; defaults and shared constraints narrow it; volume expands the candidate pool; and human filtering determines the outcome.

**3. Synthesis unconstrained by human embodiment.** Audio- and video-generation systems can synthesize performances unconstrained by human physiology, live-production conditions, or the historical prevalence of recorded styles. Music-generation systems, for example, can produce extended, breathless vocal phrases and hybrids of genres or timbres that are rare or absent in recorded corpora (57–59). Video-generation and AI-assisted editing extend the same logic to action: exaggerated gestures, impossible camera movements, fantasy transformations, stylized action sequences, and biomechanically difficult or implausible maneuvers can be generated without the same reliance on live performers, sets, stunt teams, animation, or conventional visual-effects pipelines. This shifts part of the workflow from filming and performance to generation, selection, and editing.

### *Misconception 3: "LLMs just regurgitate the training data"*

Language models can reproduce snippets of their training corpus, especially boilerplate or famous passages, raising privacy and copyright concerns (62). But treating an LLM as a giant lookup table conflates parametric compression, contextual prompting, and external storage, encouraging the assumption that any output is copied text. In practice, most generations are not copies but recombinations of patterns learned across many sources. Table 1 reframes the issue as an empirical question—when are outputs actually copied or near-verbatim?—and links mitigations to data curation, leakage auditing, and product-level memory design.

### Misconceptions about memory and alignment

#### *Misconception 4: "LLMs learn from my chats and remember me" vs. "LLMs are stateless and remember nothing"*

One pole treats a chat assistant as an ever-growing diary that learns from conversations and remembers user secrets; the other treats each reply as generated in isolation, with continuity reduced to a user-interface trick. Both neglect the three memory layers in deployed systems. Parametric memory is information encoded in the weights as a frozen, lossy compression of the training data; "stateless" here means only that those weights are fixed during inference—that is, during ordinary use (28). Contextual memory is the conversation history, retrieved content, and other input material the product includes in the current prompt, which the model can "remember" only while that material remains within the context window. External, product-level memory—logs, profiles, documents, and retrieval indices—sits outside the model and is selectively injected back into prompts; this is where many privacy and governance risks concentrate (36). Table 1 redirects the analytical focus from "does the model remember me?" to "where is this information stored, retained, and reused, and under what policy?" It also situates the relevant safeguards in explicit, auditable memory layers at each level.

#### *Misconception 5: "Fine-tuning and RLHF are just superficial filters on a neutral core"*

A common view imagines a value-neutral base model with a thin, detachable "alignment layer" on top, such that instruction tuning and RLHF simply bolt on censorship, refusals, or politeness while leaving an underlying neutral system intact. In reality, supervised fine-tuning and RLHF update the same parameters that encode the model's knowledge and skills, reshape the conditional distribution, and produce a different policy rather than a pristine core plus a mask (30). Nor is there a value-neutral baseline to recover: pretraining already reflects the norms, omissions, and sampling biases of its data, and alignment procedures further inscribe institutional preferences into behavior (1). External filters and classifiers do exist and can sometimes be added or removed without retraining; yet conflating them with RLHF encourages wishful thinking about reversibility and obscures that fine-tuning can introduce trade-offs in coverage, style, and refusal patterns. The open-weight model DeepSeek-R1 illustrates this architecture (63): censorship is implemented partly in the weights and partly through deployment-time filters (Box 2). Table 1 therefore treats fine-tuning as substantive training and asks how behavior diverges between base and fine-tuned variants when external filters are held constant.

**Box 2. DeepSeek-R1 censorship as a case study of alignment and deployment.**

DeepSeek-R1 is an open-weight model whose behavior depends heavily on post-training alignment and deployment choices (63–65).

**Training, alignment, and filters.** R1 was first pretrained, then instruction-tuned and RLHF-aligned under state-mandated constraints. In the official app, the aligned model is wrapped in server-side filters that monitor prompts and completions, interrupting or overwriting responses on politically sensitive topics (e.g. the 1989 Tiananmen protests or Taiwan's political status). The underlying weights support strong mathematical and coding performance, but political queries in that deployment yield refusals or party-line answers, and mixed prompts can cause the model to drop step-by-step reasoning once a sensitive subquestion appears.

**Open weights and "realignment."** Because the weights are publicly available, independent groups have altered R1's behavioral policy without retraining from scratch. One line of work fine-tunes R1 on factual, uncensored answers to previously suppressed questions, producing variants that retain reasoning ability while reducing political refusals. Another uses targeted weight editing or compression to identify and remove parameters most associated with censorship, reporting models that answer politically sensitive questions without refusal while largely preserving benchmark performance.

**Lessons.** Censorship in R1 is neither a thin detachable mask on a neutral core nor an immutable essence of the model. It is a substantial but partly editable component of the learned policy, supplemented by deployment-time filters. The case crystallizes three distinctions: between pretraining and post-training alignment, between behavior shaped in the weights and behavior enforced by external filters, and between one parametric system and the multiple normative regimes that can be instantiated from it. DeepSeek-R1's open weights make this decomposition more tractable; for closed-weight systems, the boundary between weight-level behavior and external filtering is empirically opaque.

## Misconceptions about cognitive status

### *Misconception 6: "LLMs think like humans" vs. "LLMs do not understand anything"*

Anthropomorphic framing treats an LLM as a unified subject with stable beliefs, goals, and feelings—an impression reinforced by interfaces and alignment procedures that encourage first-person, socially reassuring responses. The same competence-agency conflation appears in broader claims that current LLMs are already artificial general intelligence, creative in the human sense, or capable of emotions or attachments. These claims differ in content but each treats broad task performance, fluent style, creative output, or relational interaction as evidence of agency, experience, or human-like understanding (37, 66). Deflationary framing makes the mirror-image error: it treats next-token prediction and lack of embodiment as proof that the system merely manipulates symbols without access to meaning. Both views miss that current models show substantial functional understanding in bounded domains—systematic generalization, in-context learning, and multistep reasoning—while lacking embodiment, persistent online learning, and stable internal state of the kind that would support human-like, experiential understanding (28, 37, 67). Table 1 therefore treats the cognitive status misconception as a competence-agency conflation, shifting the question from "does it have a mind?" to "which tasks can it reliably perform, and what forms of agency are being implicitly attributed to it?" Describing competence functionally, rather than inferring agency or moral standing from fluent style, helps avoid both over- and under-attribution (39).

## Implications for evaluation, deployment, and governance

The taxonomy has three practical targets: capability evaluation, deployment design, and institutional governance.

### Evaluation and capability modeling

Capability evaluation should treat the deployed configuration—not the base model or a default chat sample—as the unit of analysis. This avoids two mirror-image errors: deflationary views miss closed-loop behavior, tool use, and in-context learning, whereas anthropomorphic views overgeneralize from striking demonstrations to stable competence. Evaluations should therefore specify the prompts, decoding settings, retrieval systems, tools, memory layers, and workflows under which each capability appears, test its stability under perturbation, and examine failures in realistic use.

A clinical-diagnosis example illustrates the point. In a 2024 randomized trial, physicians given access to GPT-4 scored only 2 percentage points above a conventional-resources control on diagnostic reasoning—a nonsignificant difference—even though GPT-4 alone outperformed the control group by 16 points (68). A follow-up trial using the same vignettes and scoring rubric but a redesigned collaborative workflow—independent clinician and AI assessments followed by an AI-generated synthesis—raised clinician accuracy to 82–85%, compared with 75% using conventional resources (69). The model did not change; the workflow did.

### Deployment and system design

Confusions about memory produce brittle, sometimes unsafe architectures: systems may assume continuity that the model lacks, or store and reuse user data in opaque product-level memory. Distinguishing parametric, contextual, and external memory clarifies what is fixed in weights, what can remain in short-lived context, and what belongs in explicit, auditable storage with separate access controls. Likewise, abandoning the "neutral core plus filter" myth forces designers to treat fine-tuning and RLHF as substantive, value-laden interventions whose trade-offs in coverage, style, and refusal patterns must be measured and, where possible, made reversible or at least auditable. The DeepSeek-R1 case (Box 2) shows how the same open-weight starting point can be wrapped, aligned, filtered, or realigned into divergent normative regimes depending on who controls post-training and deployment.

### Governance and public reasoning

In governance debates, treating LLMs as mere parrots understates risks arising from unreliability, scale, and misuse; treating them as proto-persons diverts debate toward premature questions of moral standing or "AI rights." The minimal model instead directs scrutiny to specific design levers: training-data curation, alignment objectives and preference-data oversight, tool and memory interfaces, and decoding and prompting defaults.

These distinctions already shape consequential legal and policy decisions, though unevenly. In *Moffatt v. Air Canada*, a tribunal held the airline liable for chatbot misinformation, rejecting the suggestion that the chatbot could bear separate responsibility; accountability attached to the operators of the deployed system. After *Mata v. Avianca* exposed fabricated AI-generated case citations, courts issued disclosure and certification orders, but some orders regulate "AI use" in the abstract rather than targeting the specific risk of fabricated sources (70, 71). The NIST Generative AI Profile offers a more architecture-sensitive approach (72): it distinguishes pretrained, adapted, and deployed systems and calls for documentation of fine-tuning, retrieval augmentation, and postdeployment monitoring—the same distinctions Fig. 1 makes explicit.

### Publisher AI policies as governance case studies

Scientific publishing provides a useful governance case study because abstract views of LLMs are translated into rules for authors, reviewers, and editors. Such policies are widespread but unevenly specified and, as currently written, limited in observed effect. Leading publishers and journals vary in what they count as "AI use" and how such use must be disclosed (73, 74). In one journal-level analysis, roughly 70% of journals had adopted AI policies, mostly disclosure requirements; nevertheless, AI-assisted writing continued to rise, explicit disclosure remained rare, and journals with policies did not differ detectably from journals without them in rates of AI-assisted writing (75).

Table 2 applies Table 1's diagnostic matrix to selected current AI policies from major publishers and psychology associations. These policies rightly emphasize disclosure, citation verification, and confidentiality, but their justificatory language also blurs the distinctions in Fig. 1. Three patterns emerge. First, passages on epistemic reliability often adopt deflationary statistical framings: they link AI-generated text or training cutoffs directly to scientific unreliability, as if the next-token objective or cutoff date determined epistemic quality by itself. This reasoning recapitulates the "just statistics" errors in misconceptions 1 and 2 by ignoring retrieval, tool use, and verification. Related warnings about reproducing large text segments echo misconception 3's lookup-table view of memorization, while giving less attention to paraphrastic or structural borrowing.

Second, passages on memory and confidentiality often treat material entered into generative AI systems as inherently accessible to providers, thereby conflating parametric, contextual, and external memory (misconception 4) and different deployment regimes. Rather than targeting data flows and data policies, they treat generative AI as a unitary data-handling regime, merging deployments with materially different storage, logging, and reuse properties into a single undifferentiated risk.

Third, statements that LLMs cannot replicate "human creative and critical thinking" embed misconception 6 in policy language by building a strong cognitive thesis into governance prose, rather than tying restrictions to task-level competence, documented reliability, and accountable use. By contrast, misconception 5 is largely absent from these policies: they say little about fine-tuning, RLHF, or how weight-level alignment interacts with external filters.

## Conclusion

Together, the minimal working model and taxonomy of six misconceptions provide a diagnostic toolkit for claims about LLMs. When evaluating claims about what these systems are, can do, or imply, the framework asks three questions: which level of description is being invoked; which distinction is being conflated; and what follows for evaluation, deployment, and governance if that distinction is kept intact. While it does not settle whether LLMs genuinely understand, reason, possess agency, or warrant moral standing, it reduces the risk of policy and practice being steered by slogans rather than by the architecture, behavior, and deployment conditions of the systems at issue.

## Acknowledgments

GPT-5.1 Pro was used to proofread the manuscript, following the prompts described at https://www.nature.com/articles/s41551-024-01185-8.

## Competing Interests

The author declares no conflicts of interest.

## Funding

The author was supported by Brain Science and Brain-like Intelligence Technology—National Science and Technology Major Project(2021ZD0204200), Yonsei University Research Fund (5355661), and the Yonsei Fellowship (funded by Lee Youn Jae).

## Author Contributions

Zhicheng Lin (Conceptualization, Visualization, Writing—original draft, Writing—review & editing)

## Data Availability

There are no data underlying this work.